\documentclass[12pt]{iopart}
\usepackage{iopams}  
\usepackage{graphicx}
\usepackage[dvips]{color}
\usepackage{cite}

\begin{document}

\title[Photoinduced magnetism and ordering in Prussian blue analog 
nanoparticles]
{Size dependence of the photoinduced magnetism and long-range 
ordering in Prussian blue analog nanoparticles of 
rubidium cobalt hexacyanoferrate}

\author{Daniel M. Pajerowski and Mark W. Meisel}
\address{Department of Physics and the Center of Condensed Matter Sciences, 
University of Florida, Gainesville, FL 32611-8440, USA}
\ead{dpaj@phys.ufl.edu and meisel@phys.ufl.edu}

\author{Franz A. Frye and Daniel R. Talham}
\address{Department of Chemistry, University of Florida, Gainesville, 
FL 32611-7200,USA}
\ead{frye@chem.ufl.edu and talham@chem.ufl.edu}

\begin{abstract}
Nanoparticles of rubidium cobalt hexacyanoferrate 
(Rb$_j$Co$_k$[Fe(CN)$_6$]$_l \cdot n$H$_2$O) were synthesized using 
different concentrations of polyvinylpyrrolidone (PVP) to produce four 
different batches of particles with characteristic diameters ranging from 3 to 13~nm.  
Upon illumination with white light at 5~K, the magnetization 
of these particles increases.  The long-range ferrimagnetic ordering 
temperatures and the coercive fields evolve with nanoparticle size.  
At 2~K, particles with diameters less than approximately 10~nm provide a 
Curie-like magnetic signal.
\end{abstract}


\section{Introduction}

The investigation of magnetism in the molecule-based solid Prussian blue,  
Fe$_4$[Fe(CN)$_6$]$_{3} \cdot x$H$_2$O, and related analogs has a rich 
history \cite{1,2}, dating back to 1928 \cite{3}.  Measurements down to 
liquid helium temperatures identified the transition to long-range 
ferromagnetic order \cite{4,5,6},  but an understanding of the 
magnetic interactions remained elusive until the 1970s, when X-ray \cite{7} 
and neutron \cite{8} diffraction data identified the crystal structure 
and the spin delocalization from high-spin Fe(III) to low-spin Fe(II).  Interest 
in these materials was renewed in the 1990s with the synthesis of several 
mixed-metal Prussian blue analogs with higher magnetic ordering temperatures,  
along with the 1996 discovery of long-lived  
photoinduced magnetism in K$_{0.2}$Co$_{1.4}$[Fe(CN)$_6$]$\cdot 6.9$H$_2$O \cite{9}.     
A flurry of experimental and theoretical research has elucidated the 
fundamental nature of the light-induced effects in three-dimensional bulk 
materials \cite{10,11}, and recent efforts have been made to integrate the 
photoinduced magnetism into films \cite{12,13,14,15,16,17,18}   
and nanoparticles, which have technical and biophysical applications.     
There have been several efforts to 
synthesize nanoparticles of Prussian blue 
analogs \cite{19,20,21,22,23,24,25,26,27,28,29,30,31}, but only a few examples 
of photoinduced 
magnetism in these particles have been reported,~\footnotemark~including 
work that isolated K$_j$Co$_k$[Fe(CN)$_6$]$_l \cdot n$H$_2$O 
particles  with typical diameters of $8 - 10$~nm  within a silica xerogel \cite{22} 
and other research producing 11~nm $\times$ 70~nm nanorods of 
Mo(CN)$_8$Cu$_2$ protected by 
polyvinylpyrrolidone (PVP) \cite{26}.  In each case,   
although photoinduced magnetism was observed, the particles did not exhibit 
long-range order.

The purpose of this paper is to report our ability to synthesize 
Rb$_j$Co$_k$[Fe(CN)$_6$]$_l \cdot n$H$_2$O nanoparticles  
protected by PVP  that exhibit photoinduced magnetism for all sizes and that 
possess long-range 
ordering, with coercive fields ranging between 0.25~$-$~1.5~kG, in the 
larger particles.  The magnetic 
properties evolve with size, where at 2~K, the particles with diameters 
less than approximately 10~nm 
provide a Curie-like magnetic signal.

\footnotetext[1]{Here, we 
differentiate photoinduced 
magnetism, which is used to describe the intrinsic process 
of the A$_j$Co$_k$[Fe(CN)$_6$]$_l \cdot n$H$_2$O Prussian blue analog, from 
photoswitchable magnetism, 
which arises from magnetization changes observed upon photoexciting a 
functionalized component or 
coating of Prussian blue analog nanoparticles, reported in Refs. 29 and 30.}

\fulltable{\label{Table1} A summary of the materials properties of the 
four sets of samples.} 
\begin{tabular}{c c c r r c c r r}
\br 
\scriptsize Batch & \scriptsize Starting & \scriptsize Resulting & \scriptsize PVP:Co & 
\scriptsize Diameter &
\scriptsize $T_{c, \mathrm{onset}}^{\,\mathrm{dark}}$
& \scriptsize $T_{c, \mathrm{onset}}^{\,\mathrm{light}}$  
& \scriptsize $H_c^{\,\mathrm{dark}}$
& \scriptsize $H_c^{\,\mathrm{light}}$\\
  &  \scriptsize PVP (g) & \scriptsize Chemical Formula~\footnotemark   &
   \scriptsize ratio   & 
  \scriptsize (nm) &  \scriptsize (K)  &  \scriptsize (K)  &  \scriptsize (G)  
  &  \scriptsize (G)\\
\mr
\scriptsize A & \scriptsize 1.0 
& \scriptsize Rb$_{1.9}$Co$_4$[Fe(CN)$_6$]$_{3.2}$ $\cdot$ ${4.8}$H$_2$O   
& \scriptsize 360 & \scriptsize $3.3 \pm 0.8$ & \scriptsize $< 2$ & \scriptsize $< 2$ 
&  \scriptsize $< 10$ 
& \scriptsize $< 10$\\
\scriptsize B & \scriptsize 0.5 
& \scriptsize Rb$_{1.8}$Co$_4$[Fe(CN)$_6$]$_{3.2}$ $\cdot$ ${4.8}$H$_2$O   
& \scriptsize 200 & \scriptsize $6.9 \pm 2.5$ & \scriptsize 10 & \scriptsize $13$ 
&  \scriptsize $\sim 15$ 
& \scriptsize $\sim 30$  \\
\scriptsize C & \scriptsize 0.2 
& \scriptsize Rb$_{1.7}$Co$_4$[Fe(CN)$_6$]$_{3.2}$ $\cdot$ ${4.8}$H$_2$O   
& \scriptsize 60 & \scriptsize $9.7 \pm 2.1$ & \scriptsize 13 & \scriptsize 17 
& \scriptsize 250 
& \scriptsize 330 \\
\scriptsize D & \scriptsize 0.1 
& \scriptsize Rb$_{0.9}$Co$_4$[Fe(CN)$_6$]$_{2.9}$ $\cdot$ ${6.6}$H$_2$O   
& \scriptsize 20 & \scriptsize $13.0 \pm 3.2$  & \scriptsize 19 & \scriptsize 22 
& \scriptsize 1000  &  
\scriptsize 1500 \\
\br
\end{tabular}
\endfulltable
\footnotetext[4]{Energy dispersive X-ray spectroscopy (EDS) on similar samples
suggests the possibility of trace amounts of K, from the K$_3$Fe(CN)$_6$ solution, 
being incorporated as interstitials.}

\section{Synthesis and Characterization}
Our Rb$_j$Co$_k$[Fe(CN)$_6$]$_l \cdot n$H$_2$O nanoparticles were synthesized by 
modifying the procedure for Prussian 
blue nanoparticles described by Uemura and coworkers \cite{23,24}.  A 2~mL solution 
containing both 28~mg K$_3$Fe(CN)$_6$ (0.085~mM) and 6.8~mg RbNO$_3$ (0.046~mM) was added 
dropwise to an 8~mL solution containing 30~mg Co(NO$_3$)$_2 \cdot 6$H$_2$O (0.103~mM) 
and PVP while stirring rapidly.  By varying the amount 
of PVP (Table~1), the protocol produced samples with different particle sizes 
and size distributions.  
After 30 minutes of stirring, the solution was allowed to sit for one week.  
For transmission electron 
microscopy (TEM) studies, a 50~$\mu$L aliquot of the suspension was diluted 2000~times, 
and 8~$\mu$L of the 
diluted suspension was placed on a holey carbon grid.  A representative image 
is shown in Fig.~1, and 
selected area electron diffraction was compared to powder X-ray diffraction 
patterns to confirm the 
structure \cite{33}. In addition, infrared spectra displayed a C$-$N stretch of 
2124~cm$^{-1}$, 
consistent with a Co(III)$-$Fe(II) low-spin pair.  Using Image J imaging 
software \cite{34}, 
the TEM images were analyzed to obtain the 
particle size distributions shown in Fig.~2.~\footnotemark~These data 
were fit to a log-normal 
function that yielded 
the characteristic diameters shown in Fig.~2 and Table~1.  To isolate the 
particles, three volumes of 
acetone were added to the synthesis solution, which was centrifuged, and 
then further washed with 
acetone and dried under vacuum.  Chemical analysis was obtained from a 
combination of CHN 
combustion analysis and inductively coupled plasma mass spectrometry (ICP-MS), 
and the resulting 
formulae are listed in Table~1, along with the ratio of the PVP repeat unit per 
cobalt.  The extent of 
H$_2$O coordination to the Co was estimated by considering the Fe vacancies 
implicit with the 
measured Co:Fe ratio. 

\footnotetext[2] {Similar size distributions have been 
obtained for Prussian blue nanoparticles protected by PVP \cite{28}.}  

\begin{figure}
\begin{center}
\includegraphics[width=3.00in]{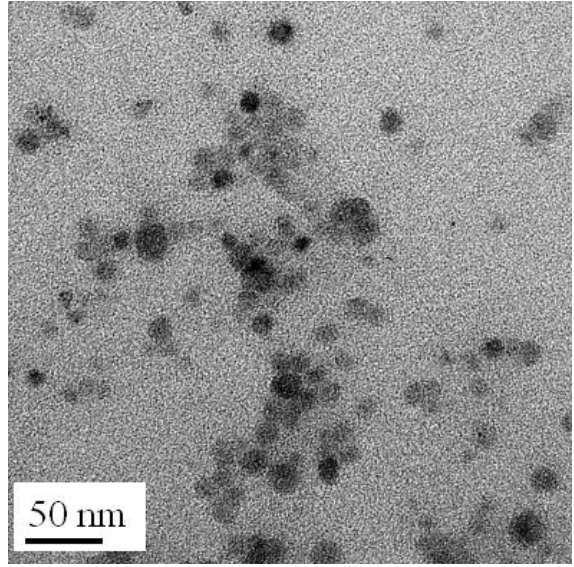}
\caption{TEM image of a collection of 
Rb$_{0.9}$Co$_4$[Fe(CN)$_6$]$_{2.9} \cdot $(H$_2$O)$_{6.6}$ 
nanoparticles from Batch~D, see Table~1. \label{Fig1}}
\end{center}
\end{figure}

\begin{figure}
\begin{center}
\includegraphics[height=4.50in]{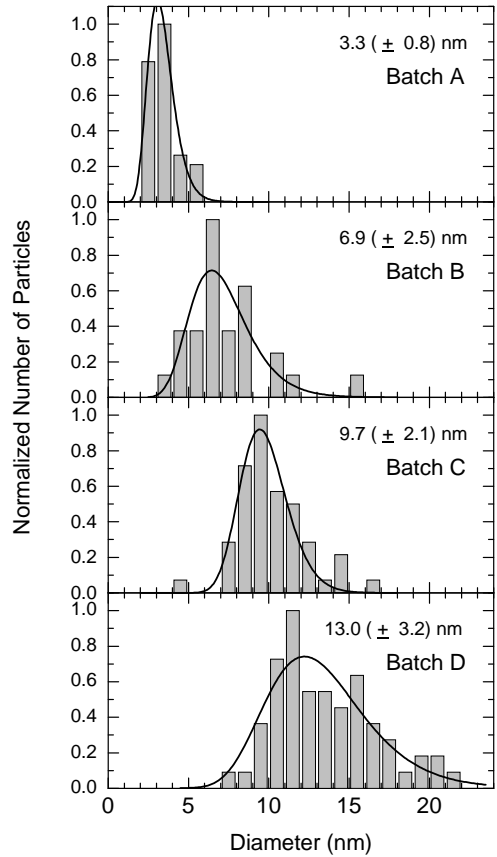}
\caption{The particle distributions, normalized to the largest bin,
versus diameter for the four batches of particles, see Table~1.
The total number of particles for
each distribution, smallest to largest, is 44, 27, 53, and 62,
respectively.  The solid lines are the results of log-normal fits
that provide the characteristic diameters shown for each
distribution.  \label{Fig2}}
\end{center}
\end{figure}

\section{Photoinduced Magnetism}
The temperature dependences of the dc magnetic susceptibilities, $\chi(T) = M/H$, 
of the four batches of particles 
are shown in Fig.~3.  A standard commercial SQUID magnetometer was employed, 
and the samples 
were mounted to commercial transparent tape and could be irradiated with 
light from a room-temperature, 
halogen source by using a homemade probe equipped with a bundle 
of optical fibers \cite{17}.    
Small diamagnetic background contributions from the holder and tape 
(representing $\lesssim 0.1$\% of the total signal) 
were independently measured and have been 
subtracted from the data.  The magnetic signals are expressed per mole of 
sample using the chemical 
formula listed in Table~1.  The dark state ZFC data were obtained after 
cooling in zero applied field 
from 300~K, while the dark state FC data were taken after cooling in 
100~G from 300~K.  The light state 
was established after field cooling the samples from 300~K to 5~K in 
100~G and subsequently irradiating 
with light for 5~hours, which saturated the photoinduced response.  
In order to avoid spin glass-like relaxation, the light state FC data were 
obtained after cycling the sample to 30~K in 100~G. 

\begin{figure}
\begin{center}
\includegraphics[width=3.00in]{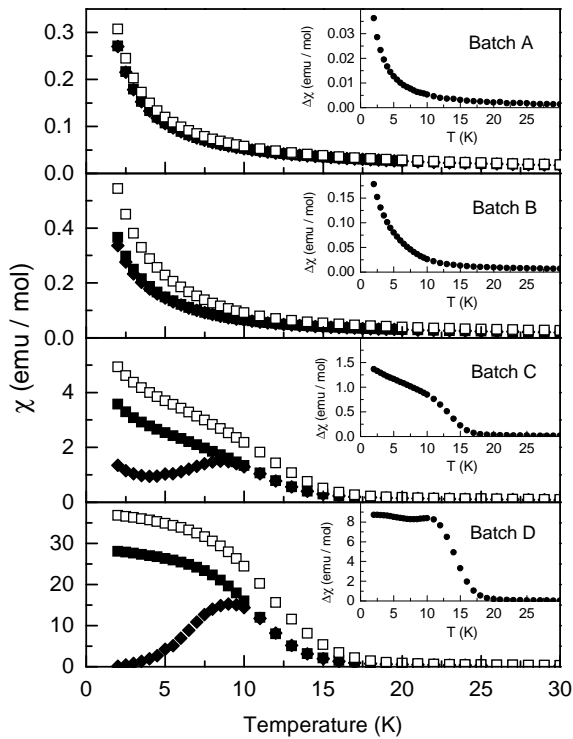}
\caption{The temperature dependences of the low field, 100~G,
susceptibilities are shown for the zero-field cooled (ZFC) dark
($\blacklozenge$), field-cooled (FC) dark ($\fullsquare$), and FC
light ($\opensquare$) states of each batch produced. The insets display
the differences between the FC light and dark states, as described
in the text.  Finite values for this difference can only arise from
photoinduced magnetism. \label{Fig3}}
\end{center}
\end{figure}

\section{Discussion}
It is noteworthy to recall that the photoinduced magnetism of the 
Prussian blue analogs depends upon 
a charge transfer induced spin crossover event (CTIST effect), 
requiring the presence of vacancies to 
allow crystalline flexibility \cite{36,37,38,39,40,41,42,43}.  
More specifically for Rb$_j$Co$_k$[Fe(CN)$_6$]$_l \cdot n$H$_2$O, 
the spins of an Co$-$Fe pair can exist in either of two arrangements.  
The low-spin state consists of Co(III) ($S = 0$) and Fe(II) ($S = 0$), 
while the high-spin state possesses Co(II) ($S = 3/2$) and Fe(III) ($S = 1/2$).  
The presence of the CTIST effect depends on the local chemical 
environment defined by the values of $j$, $k$, $l$ and $n$ in the 
chemical formula.  For limiting ratios, Co and Fe spins can be locked 
into either their high-spin or low-spin states for all accessible temperatures.  
Alternatively, with the proper tuning of the chemical ratios, 
the Co and Fe ions can exist in high-spin states at room temperature 
and then experience a crossover to their low-spin states at approximately 150~K.  
The spin crossover phenomenon generates the pairs involved in the photoinduced magnetism.  
However, since the CTIST effect may not be 100\% efficient, some regions 
remain locked in their high-spin states, and we refer to them as 
{\it primordial} spins \cite{13}.  For example, these primordial spins are 
responsible for the magnetic signals observed for the dark state of 
the largest particles.  When irradiated, the low-spin regions, 
near the primordial high-spin clusters, are 
photoinduced to the high-spin magnetic state, resulting in a growth 
of the magnetic domain.  This scenario is supported by our description of 
the anisotropic photoinduced magnetism observed in thin 
films \cite{13,18} and by the local probe investigations of others \cite{45,46}.  
All samples show a photoinduced increase in their magnetic signals, 
and the strength of the change is correlated with the size of the particles.  
The differences between the FC susceptibilities of the light and dark states, 
\begin{equation}
\Delta \chi = \chi^{light}_{FC} - \chi^{dark}_{FC} \;\;\;,
\end{equation} 
are plotted in the insets of Fig.~3, and finite values can only arise from the 
photoinduced magnetism.  
It is important to stress that photoinduced magnetism is observed even in small 
particles possessing only Curie-like behavior.

\begin{figure}
\begin{center}
\includegraphics[width=3.00in]{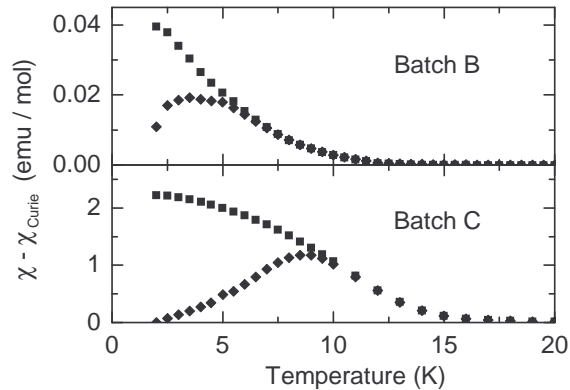}
\caption{The dark state, i.e.~primordial, ZFC ($\blacklozenge$) 
and FC ($\fullsquare$) states show a signal associated with long-range 
order for the two batches displaying a mix of magnetic behavior 
after subtracting the Curie-like contribution.  The susceptibility 
of Batch~A only exhibits a Curie-like contribution, while the magnetic 
signal for Batch~D is dominated by spins that are ordered, see text for details.  
For each sample, a Curie-like contribution was obtained by fitting the 
dark state data above 30~K, and this magnetic signal was 
subtracted from the FC and ZFC dark state data shown in Fig.~3.
\label{Fig4}}
\end{center}
\end{figure}

Two main features are seen when considering the evolution of 
material properties due to the 
increasing average size of the separate batches, 
namely the onset of long-range magnetic order and an 
increasing net magnetization.  This scaling of magnetization 
is linked to an increased diamagnetic-surface to magnetically-active-volume 
ratio at smaller particle sizes.  At low temperatures, the high-spin 
Co and Fe ions interact antiferromagnetically, giving rise to a 
ferrimagnetic transition at $T_C$, which is about 24~K for bulk powder 
specimens with a similar chemical formula \cite{38}.  For the magnetic data shown 
in Fig.~3, the onset of this transition can be estimated, 
and these macroscopic temperatures are listed in Table~1.  
We contend that particles larger than a critical size will 
allow domains large enough to approach bulk-like magnetic properties.  
Conversely, smaller particles may put limits on allowed 
domain size, suppressing the ordering temperature.  Microscopically, 
if the size of the magnetic domains is less than or of the order of the 
magnetic coherence length, then a spectrum of $T_C$ values can be 
expected, and evidence of this trend is observed in Batches C and D, 
see the insets of Fig.~3.  In other words, the size distributions  
may include particles both above and below the critical size 
that leads to bulk-like properties.  As a result, the magnetic response 
of each batch may include contributions from Curie-like and bulk-like fractions.  
The percentage of each can be estimated by subtracting a Curie-like 
signal from each $\chi(T)$ plot, and the results for Batches B and C 
are shown in Fig.~4.  In contrast, Batch~A follows Curie-like behavior, 
whereas the active sites in Batch~D are almost 
entirely ferrimagnetically ordered.  This analysis of the data 
suggests the Curie-like contribution for each 
of the four batches of nanoparticles (smallest to largest) is 
100\%, 90\%, 50\%, and 10\%.  Consequently, at least down to 2~K, 
nanoparticles with sizes below $\sim 10$~nm do not exhibit long-range order.  
These interpretations are consistent with our $M$ versus $H$ measurements 
performed at 2~K, where coercive fields, $H_C$, and remnant 
magnetization values are observed for the largest sets of particles but not for the 
smallest set of particles, Fig.~5 and Table~1.  Furthermore, 
the differences between the FC and ZFC data for the dark states in 
Batches B, C, and D are consistent with spin glass or 
cluster glass behavior \cite{47,48}, and our ac susceptibility data are 
presented in the Appendix.

\begin{figure}
\begin{center}
\includegraphics[width=5.00in]{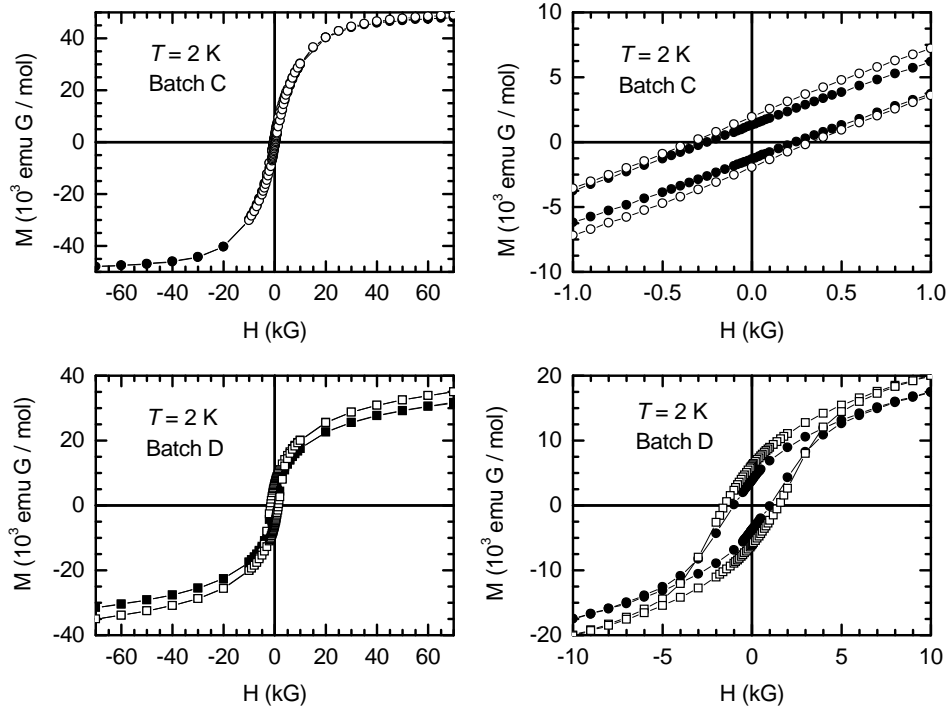}
\caption{The $T = 2$~K magnetization versus magnetic field 
sweeps for the two largest sizes of nanoparticles, Batches C and D, 
are shown for the light ($\opencircle$ and $\opensquare$) 
and dark ($\fullcircle$ and $\fullsquare$) states.  The coercive fields, 
$H_C$ for the light and dark states for each batch are listed in Table~1, 
and the lines are guides for the eyes. 
\label{Fig5}}
\end{center}
\end{figure}

\section{Conclusions}
In conclusion, we have synthesized four different sizes of 
Rb$_j$Co$_k$[Fe(CN)$_6$]$l \cdot n$H$_2$O 
nanoparticles protected by PVP.  Each batch of particles exhibits photoinduced magnetism, but 
the strength of this effect, as well as other global properties, e.g.~$T_C$ and $H_C$, 
are correlated with the intrinsic particle size distributions of each batch.  The combination of 
photoinduced magnetism and nanosized Prussian blue 
analog particles with finite coercive fields is unique and establishes a length scale limit of 
$\sim 10$~nm for these properties.  The next advances in the field will undoubtedly employ Prussian 
blue analogs with higher transition temperatures \cite{49} 
that may have a variety of technical applications, 
including, for example, the exploitation of optically controlled 
magnetocaloric properties \cite{50}.

\ack{This work was supported, in part, by NSF DMR-0305371 (MWM) and NSF 
DMR-0543362 (DRT).  We acknowledge early contributions by N.~E.~Anderson,  J.~Long and 
J.-H.~Park.  We thank the UF Department of Geology for the ICP-MS work and the UF Major Analytical 
Instrumentation Center for the TEM images.}

\appendix
\section*{Appendix}
\setcounter{section}{1}
\begin{figure}
\begin{center}
\includegraphics[width=5.00in]{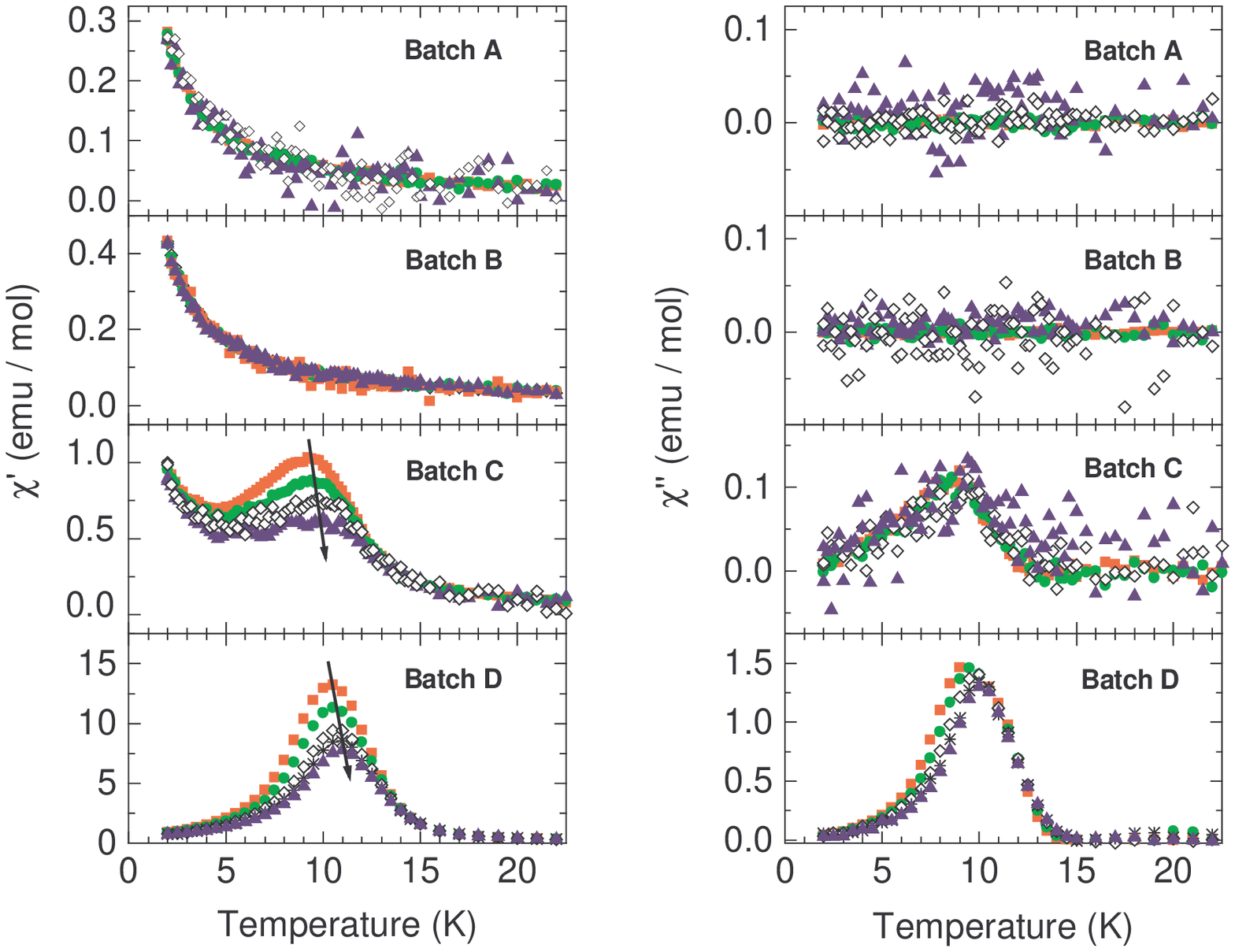}
\caption{The temperature dependences of the real ($\chi^{\prime}$) and imaginary 
($\chi^{\prime \prime}$) ac susceptibilities of the ZFC dark states, i.e.~primordial 
states, are shown for all four batches of nanoparticles.  All batches 
were measured with no applied static field and an alternating field of 4~G, 
except for Batch~D, which was measured in 1~G.  The frequency dependence was 
studied at 1~Hz (\textcolor{red}{$\fullsquare$}), 10~Hz
(\textcolor{green}{$\fullcircle$}), 100~Hz ({$\Diamond$}), and 1~kHz 
(\textcolor{blue}{$\fulltriangle$}) for all 
batches, except for Batch~D, which has an additional measurement at 
333~Hz~($\ast$).  Arrows are guides for the eyes and pass through the peaks.}
\end{center}
\end{figure}
The temperature dependences of the real ($\chi^{\prime}$) and imaginary 
($\chi^{\prime \prime}$) ac susceptibilities of the ZFC dark states, i.e.~primordial 
states, of all four batches are shown in Fig.~6.  The phenomenological 
parameter, $\beta$, given by \cite{47}
\begin{equation}
\beta\;=\;\frac{\Delta T_f}{T_f \Delta(\log \omega)} \;\;\;,
\end{equation}
where $T_f$ is the freezing temperature given by the cusp in $\chi^{\prime}(T)$ and 
$\omega$ is the angular frequency, is $0.024 \pm 0.004$ for Batches C and D, 
and this observation is consistent with spin glass or cluster glass 
behavior \cite{47,48}.

\end{document}